# An Optical Method for Evaluating the Mechanical Properties of Wires Under Impact Tensile Load

Phichai Youplao, Hadi Nasbey, Akihiro Takita, Prin Nachaisit, and Yusaku Fujii

*Abstract*—The dynamic properties of materials utilized in architecture or engineering applications can significantly affect their performance under dynamic or impact loading conditions. To evaluate such behavior, force transducers are commonly employed in testing. However, the calibration of force transducers is typically limited to static conditions and relies solely on gravitational forces exerted on standard masses. Thus, assessing the uncertainty in force measurements using force transducers during dynamic loading conditions remains a challenging task, presenting a significant obstacle in accurately characterizing the dynamic behavior of materials. In this work, an optical technique to evaluate the mechanical properties of wires subjected to impact tensile loads is presented. The wire under test is subjected to an impact tensile load by applying the inertial force of a rigid mass, which is supported by utilizing an aerostatic linear bearing with sufficiently small friction. The inertial force applied to the wire can be determined by multiplying the mass of the rigid mass by its acceleration, where the acceleration can be measured employing a Michelson type optical interferometer. The performance of the proposed method is demonstrated through experiments and analysis of the dynamic characteristics of a tungsten wire under impact tensile loading conditions.

*Index Terms*—Definition of force, deformation, dynamic load, elasticity, inertial force, inertial mass, interferometry, tensile stress.

## I. INTRODUCTION

The demand for accurately assessing and predicting the mechanical properties of materials and structures when subjected to dynamic or varying loads is increasingly important for designing and optimizing complex systems in various industrial and research contexts. This need has become apparent in several domains, including material testing and modal analysis [1-6]. In order to gain a more accurate and reliable understanding of the dynamic behavior of materials, numerous studies have been carried out due to the fact that, in actual usage conditions, the materials are constantly subjected to dynamic loads in various environments. For instance, an experimental method involved solving Newton's second law equation under dynamic stress [7], to measure the dynamic elastic modulus of several polymeric soft materials. A study aims to assess the capacity of the substructure to withstand both static and dynamic loads generated by a hammer machine [8], utilizing a high-strain dynamic load test to supplement the limited information obtained from the conventional static load test. The dynamic behavior of butt welds and fillet welds in steel structures was studied using compression-tension conversion tests [9], and results showed that increasing strain rate increases strength, but affects fracture strain and failure modes differently based on weld strength matching. Furthermore, research into the dynamic properties of various interesting materials, such as organic compounds [10], coal [11], reinforced cementitious composites [12], and glass [13], has been carried out with a focus on specific applications. In such applications, force transducers are commonly utilized for measuring the varying force. However, force transducers are usually calibrated only under static conditions employing static forces, which are based on the gravitational forces acting on the standard masses. At this moment, there are no standardized methods for dynamic force calibration. This makes it challenging to evaluate the uncertainty in a measured force when implementing a force transducer under dynamic conditions.

To overcome this situation, the Levitation Mass Method (LMM), which generates and measures a varying/dynamic force as the inertial force of a mass, has been proposed and developed by the authors. The LMM utilizes the inertial force of a mass that is levitated using a pneumatic linear bearing as the reference force to apply to the objects under test [14]. These objects can include force transducers [15], materials [16], or structures [17]. The collision between the mass and the objects being tested under impact compressive load allows for the evaluation of the mechanical properties of these objects. An optical interferometer is implemented to measure the inertial force of the levitated mass during the collision. During the measurement process, only the motion-induced time-varying beat frequency, which is obtained from the Doppler shift frequency of the laser light reflected on the mass, is measured in the LMM. All other quantities, including velocity, position, acceleration, and force, are then numerically calculated from the frequency. As a result, the obtained quantities are in good synchronization with each other. Additionally, force is directly calculated based on its definition, which is the product of mass and acceleration.

Tungsten is a dense, silvery-white metal with a high melting point, and is highly resistant to corrosion and wear [18, 19]. These unique properties make it an important material for a wide range of various applications in industries, including the steel industry, used as alloying elements to improve the strength and toughness of steel [20, 21], and military applications. In the electronics industry, tungsten is used to manufacture electrical components and devices as well, such as filaments for light bulbs and high-speed electrical contacts [22], sensors [23], and electronic devices [24]. However, to enhance the dependable use of tungsten across industries, the examination of its mechanical properties and mechanisms of deformation under impact tensile load conditions, with an appropriate method for dynamic force calibration, is essential.

In this work, a novel approach is presented to evaluate the dynamic mechanical characteristics of tungsten wires subjected to impact tensile loading conditions, by implementing the LMM. The experimental results obtained using the proposed method, along with an approximating calculation to estimate the value of the impact tensile stress, have undergone validation and analysis, which will be discussed in this article. The findings of this study have significant implications for the design and optimization of materials and structures that are subject to dynamic loading conditions.

## II. EXPERIMENT

The experimental setup used to measure the mechanical properties of wires subjected to impact tensile load is illustrated in Fig. 1. A tungsten wire with a diameter of 0.1 mm (model: W-461167, 99.95% purified, manufacturer: The Nilaco Corp., Japan) was used as the test specimen. The wire was fixed in place using two clamps, and a photograph of the test section is provided in Fig. 2. The natural length of the wire under test is set to 100.0 mm.

To enable negligible friction in the horizontal linear motion of the mass during impact tensile testing, an aerostatic linear bearing (Model: VAPB510Y, manufacturer: TOTO Co., Ltd., Japan) was utilized. The moving part (the considered mass) of the bearing was adjusted so that its center of gravity aligned with the two clamp points of the wire, by adding additional masses. The total mass of the whole moving mass components, including a cube corner prism, a clamp, and some additional masses, was set to m = 2.897 kg. A Zeeman-type He-Ne Laser (model: HP-5517C, manufacturer: Agilent Technologies, Inc., USA), of which the laser beam exhibited two frequencies in orthogonal polarizations, was utilized as the light source for the optical interferometer. The interferometer was securely installed on a standard honeycomb stainless-steel optical table, enabling control of both its balance and any vibrations that may impact the system.

To initiate the measurement, the wire prepared to begin at a loose position. Next, an initial velocity is manually given to the moving mass, causing it to move in the right direction as shown in Fig. 1. When the wire becomes taut, a pair of action and reaction forces applies an impact tensile force to both the wire and the moving mass. The value of the impact tensile force applied on the wire, $F$, can be determined by measuring the inertial force acting on the moving mass, which is calculated by multiplying the mass of the whole moving mass components by its acceleration.

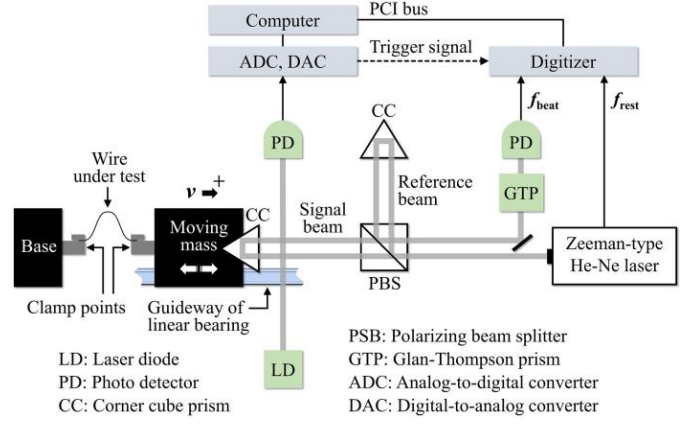

Fig. 1. The experimental setup.

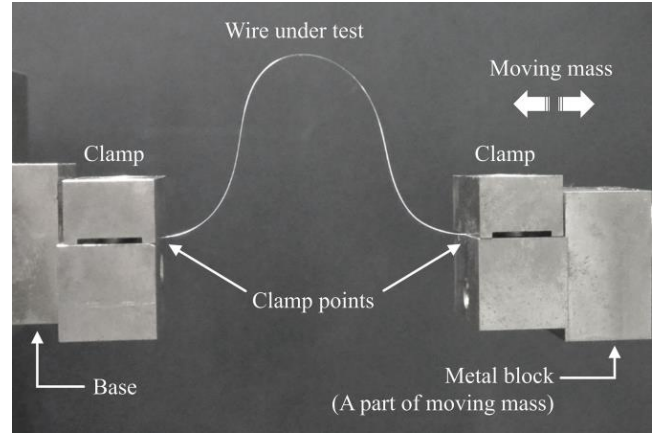

Fig. 2. Photograph of the test section.

The acceleration, $a$, is derived by differentiating the velocity-time history of the mass, where the velocity, $v$, of the moving mass is calculated from the Doppler shift frequency of the signal beam obtained by the interferometer, $f_{Doppler}$, which can be expressed as:

$$v = \lambda_{air}(f_{Doppler})/2 \quad (1)$$

$$f_{Doppler} = -(f_{beat} - f_{rest}) \quad (2)$$

The wavelength of the signal beam in air is $\lambda_{air}$ = 632.8 nm. The beat frequency, represented by $f_{beat}$, corresponds to the disparity in frequency between the signal beam and the reference beam, which can be measured as the frequency of the electrical signal output from the photo diode (PD) of the optical interferometer. The rest frequency, represented by $f_{rest}$, corresponds to the difference between the two frequencies in orthogonal polarizations of the emitted laser light, which is approximately 3.13 MHz. The rest frequency, $f_{rest}$, can be measured as the frequency of the electric signal provided from the photo diode embedded in the laser head. When the moving mass is stationary, which resulting in the absence of a Doppler shift in the signal beam, the beat frequency, $f_{beat}$, and the rest frequency, $f_{rest}$, are equivalent.

The electrical signals obtained from the photo diodes are digitized utilizing a high-speed digitizer (model: NI PCI-5105, manufacturer: National Instrument Corp., USA) that samples each channel at a rate of 30 M samples/second with a resolution of 8 bits. This setup allows to obtain the data samples up to 15 M samples per channel for a sampling duration of approximately 0.5 seconds. Determination of the beat and the rest frequencies, $f_{beat}$ and $f_{rest}$, from the digitized waveforms can be achieved accurately employing the Zero-crossing Fitting Method (ZFM) [25]. In ZFM, the frequency is determined by considering all the zero-crossing times within a defined sampling interval of the waveform. In this investigation, the sampling interval was defined as N = 2000 periods, which is corresponding to approximately 0.64 ms when the frequency is approximately of 3.13 MHz.

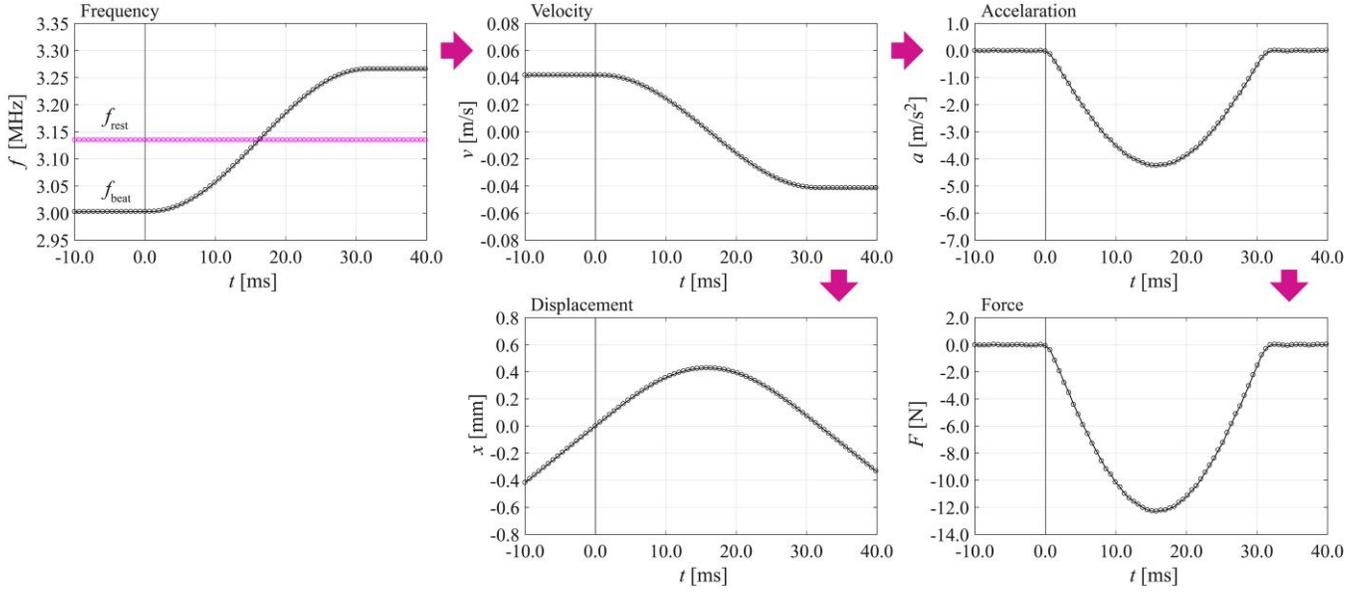

Fig. 3. Data processing procedure.

The digitizer is activated by a trigger signal generated from a digital-to-analog converter (DAC). This stimulus is initiated when the movement of the mass obstructs the light passing through a laser diode and a photodiode, which work together as a trigger switch.

Thirty sets of impact tensile force measurements were performed in the experiment with varying initial velocities of the moving mass.

III. RESULTS

Fig. 3 illustrates the data processing procedure for one of the 30 measurements. The frequencies $f_{beat}$ and $f_{rest}$ are utilized to calculate the velocity $v$, the displacement $x$, the acceleration $a$, and the force $F$.

In Fig. 3, an estimated maximum value of the impact tensile force, $F_{max}$, of approximately -12.30 N, along with an estimated temporal full width at half maximum, FWHM, of the impact tensile force, $T_{FWHM}$, of approximately 19.8 ms, are revealed. To determine the origin of the time and displacement axes, the following procedures were followed: firstly, the threshold value was established using 0.1% of the estimated maximum impact tensile force, $F_{max}$. Secondly, the impact tensile force data was traced back to the first data point from the maximum impact tensile force. Finally, the corresponding time and displacement of the traced data of impact tensile force that is less than the value of the threshold and greater than zero were designated as the origin time and origin displacement.

Fig. 4 presents the results of the impact tensile measurement described in Fig. 3, including a plot of the change in velocity as a function of displacement in Fig. 4(a). The kinetic energy dissipated during the measurement is calculated using the velocities before and after the tensile force is applied to the wire, which are approximately $v_1 = 4.18 \times 10^{-2}$ m/s, and $v_2 = -4.14 \times 10^{-2}$ m/s, respectively. Based on these values, the kinetic energy dissipation is estimated to be approximately $4.82 \times 10^{-5}$ J.

Fig. 4(b) shows the correlation between the change in strain rate and the strain in the tungsten wire under investigation. During the initial stages of deformation (represented by the black line), the correlation was likely to remain linear with a gradual decrease in strain rate values, from 0.4177 magnitude/s to 0.3849 magnitude/s, when the strain did not exceed 0.2%. Subsequently, as the tungsten wire approached the maximum exerted stress, the strain rate then decreased rapidly. The maximum strain, Strain$_{max}$, observed was approximately 0.4283% at an estimated strain rate of 0.0195 magnitude/s. Once the stress was relieved, the tungsten wire regained its original shape, exhibiting behavior that closely resembled its initial elasticity (represented by the magenta line).

The relationship between the force acting on the moving mass and the displacement is depicted in Fig. 4(c). The spring constant decreases as the displacement increases and eventually becomes almost constant when the displacement exceeds 0.1 mm. According to the Newton's Third Law, the force from the moving mass that acting on the tungsten wire is expressed as $-F$. The maximum tensile impact force, $F_{max}$, was observed to be approximately -12.30 N at a maximum displacement, $x_{max}$, of approximately 0.4283 mm. The work done by the moving mass was computed by integrating the force over the motion trajectory and was estimated to be approximately $4.75 \times 10^{-5}$ J. This value is consistent with the dissipation of kinetic energy.

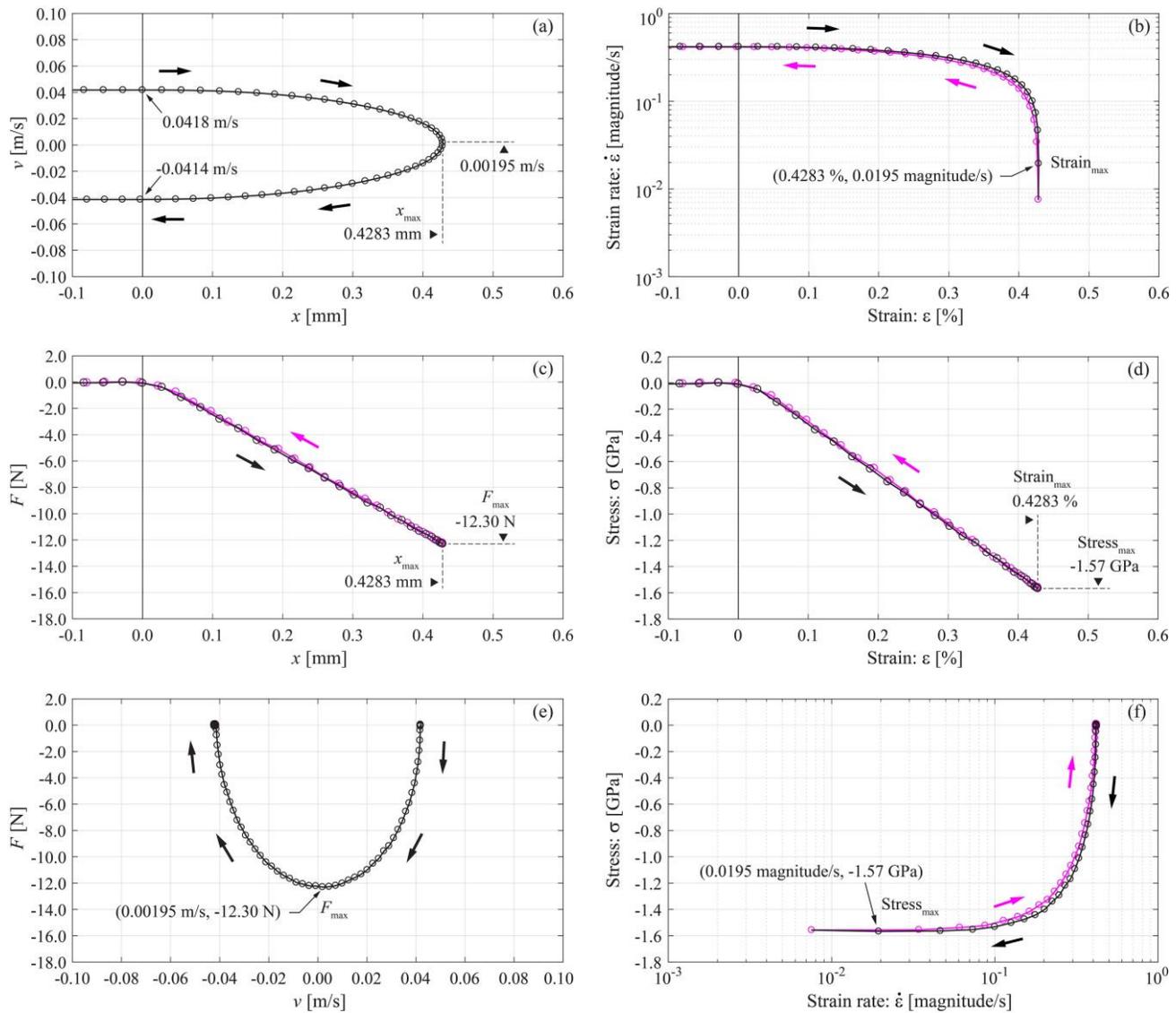

Fig. 4. Comparison results, (a) Variation in velocity correlated to displacement, (b) Variation in Strain rate correlated to Strain, (c) Variation in force correlated to displacement, (d) Variation in Stress correlated to Strain, (e) Variation in force correlated to velocity, and (f) Variation in stress correlated to strain rate.

Fig. 4(d) illustrates the stress-strain relationship of the tungsten wire under test. In the early stage (black line), the wire received stress and stretched, a linear relationship was observed until reaching the maximum stress of approximately -1.57 GPa at the maximum strain of approximately 0.4283%. Afterward, when the stress was relieved, the tungsten wire regained its original shape and manifested behavior that closely resembled its initial elasticity (magenta line).

The relationship between the force acting on the moving mass and the velocity is illustrated in Fig. 4(e), where a nearly symmetrical curve is observed due to the small amount of kinetic energy dissipation during the measurement. The velocity, $v$, at which the moving mass was experiencing the maximum tensile impact force, $F_{max}$, of approximately -12.3 N, was estimated to be $1.95 \times 10^{-3}$ m/s

Fig. 4(f) illustrates the relationship between stress and strain rate of the tungsten wire under test, which is in agreement with the correlation demonstrated in Fig. 4(b). During the initial stages of deformation (black line), the stress and strain rate relationship was likely to remain linear and observed a rapid increase in stress values up to -0.56 GPa during the strain rate was approximately greater than 0.4000 magnitude/s. However, as the tungsten wire approached a minimum strain rate, the stress increased gradually. The maximum stress, Stress$_{max}$, was approximately -1.57 GPa at the strain rate of approximately 0.0195 magnitude/s. Upon release of stress (magenta line), the tungsten wire returned to its original shape, exhibiting behavior that closely resembled its initial elasticity.

The strain energy, $U$, which is equal to the work done by the moving mass and represents the elastic potential energy accumulated by the wire as a result of recoverable elastic deformation within the stress-strain curve, can be expressed as:

$$U = \frac{1}{2}Fx = \frac{1}{2}V\sigma\varepsilon \qquad (3)$$

where $F$ is the impact tensile force, $x$ is the displacement, $V$ is volume of the tungsten wire body, $\sigma$ is stress, and $\varepsilon$ is strain. The correlation between strain energy and strain can be depicted in Fig. 5. The strain energy is in concurrence with the result obtained by integrating the force along the displacement as shown in Fig. 4(c), which is equivalent to integrating the stress along the strain as shown in Fig. 4(d) multiplied by $V$. For example, the maximum strain energy, $U_{max}$, was calculated to be approximately -2.63 mJ at the maximum strain of approximately 0.4283%. The work done by integrating the force along the displacement, as shown in Fig. 4(c), and by integrating the stress along the strain, as shown in Fig. 4(d) multiplied by $V$, was calculated to be approximately -2.54 mJ.

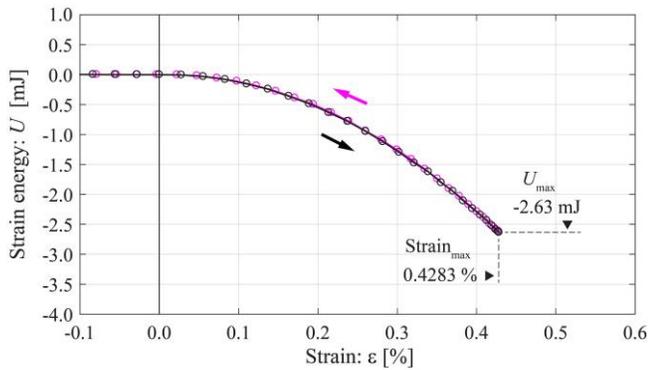

Fig. 5. Variation in strain energy as a function of strain.

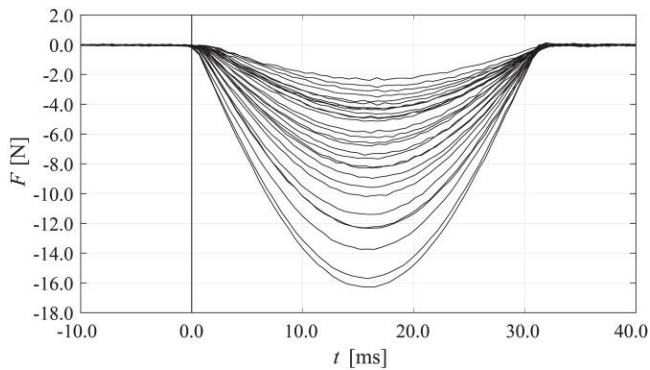

Fig. 6. Variation in impact tensile force as a function of time for the 30 measurements.

The variation in the force, $F$, with respect to time, $t$, was investigated through the analysis of 30 measurements, of which the results are depicted in Fig. 6. The data reveal a consistent direction between the impact tensile force and time. The maximum impact tensile force, $F_{max}$, was found to vary within the range of approximately -2.40 N to -16.28 N.

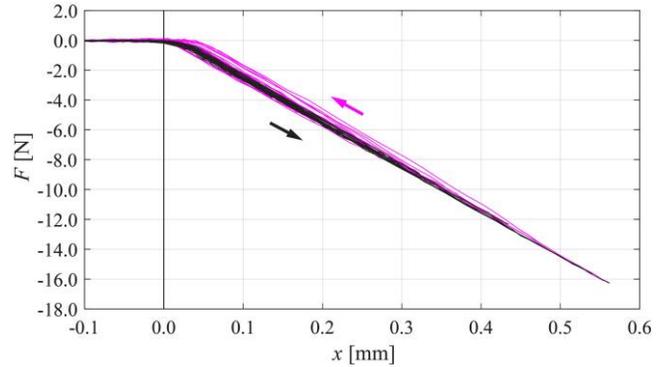

Fig. 7. Variation in impact tensile force as a function of displacement for the 30 measurements.

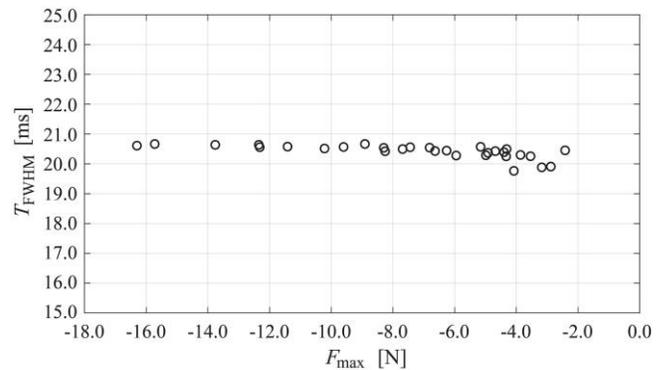

Fig. 8. The correlation between FWHM values of the measured forces $F$ and the corresponding $F_{max}$ for the 30 measurements.

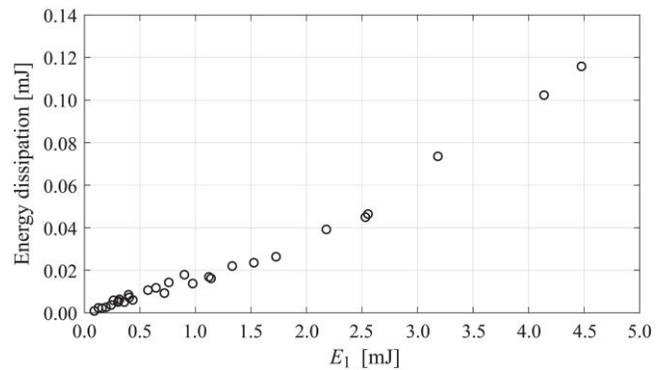

Fig. 9. The correlation between the energy dissipation ratio and the initial kinetic energy of the moving mass for the 30 measurements.

Fig. 7 illustrates the relationship between the force, $F$, and displacement, $x$, for the 30 impact tensile force measurements. This trend is comparable to that observed in Fig. 4(c). The maximum impact tensile force, $F_{max}$, as shown in Fig. 6, corresponded to a maximum displacement, $x_{max}$, range of approximately 0.0853 mm to 0.5619 mm, respectively. Furthermore, the initial point of the impact tensile force was consistently detected at the beginning of each measurement, and the linear portions of all the curves aligned well. Consequently, it can be inferred that all the curves exhibit equivalent spring constants, indicating the reproducibility of the proposed method.

Fig. 8 shows the correlation between the temporal full width at half maximum (FWHM), $T_{FWHM}$, and the maximum value of the impact tensile force, $F_{max}$, obtained through 30 impact tensile force measurements. The spring constant of the tungsten wire remains nearly constant with an increase in $F_{max}$, evidenced by the consistent FWHM of the impact tensile force relative to the increase in $F_{max}$.

Fig. 9 shows the correlation between the kinetic energy dissipation and the initial kinetic energy of the moving mass, as determined from 30 impact tensile force measurements. The kinetic energy dissipation increases as the initial kinetic energy increases, where the maximum kinetic energy dissipation was observed to be approximately 0.116 mJ at the initial kinetic energy of approximately 4.480 mJ.

## IV. Discussion

### A. Approximation of impact tensile stress

A relatively simple model, compared to other more complex models, and widely used to describe the mechanical behavior of materials that exhibit elastic and viscous properties, known as the Kelvin-Voigt model, is satisfactorily employed to perform an approximation of the impact tensile stress in this investigation. The representative diagram of the Kelvin-Voigt model is thus placed instead of the tungsten wire as illustrated in Fig. 10.

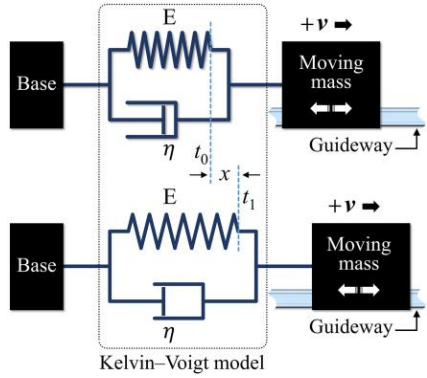

Fig. 10. Schematic depiction of the Kelvin–Voigt model.

In accordance with the model definition, the two components (spring and dashpot) are connected in parallel, resulting in equal strains in each component, which can be expressed as:

$$\varepsilon_r = \varepsilon_s = \varepsilon_d \quad (4)$$

where $\varepsilon_r$, $\varepsilon_s$, and $\varepsilon_d$, are the resultant strain, the strain in the spring component, and the strain in the dashpot component, respectively. Therefore, the same resultant stress is applied to both components at the same time and can be calculated as the sum of the stress in each component. The resultant stress can be mathematically represented by the linear first-order differential equation, expressed as:

$$\sigma_{cal}(t) = c + E\varepsilon(t) + \eta \frac{d\varepsilon(t)}{dt} \quad (5)$$

where $\sigma_{cal}(t)$ is the calculated stress, $\varepsilon(t)$ is the strain, $d\varepsilon(t)/dt$ is the time derivative of the strain (or strain rate: $\dot{\varepsilon}$), all at time $t$, $c$ is a constant coefficient, $E$ is the elastic modulus of the material, and $\eta$ is the viscosity coefficient of the material.

In this study, the coefficients, $c$, $\eta$, and the elastic modulus of the tungsten wire, $E$, were estimated utilizing multiple least squares regression analysis, with specifically selected data during the moments that the wire was experiencing the dynamic impact loading from the 30 experiments. Therefore, to achieve the estimated values of the considered quantities, a force was manually applied to the moving mass at varying initial velocities in the experiments. During each trial, data-time history profiles of the resultant force and the displacement were collected and employed to calculate the corresponding stress, $\sigma_{mea}$, strain, $\varepsilon$, and strain rate, $\dot{\varepsilon}$. These data-time history profiles with 1,544 samples were then analyzed using multiple regression analysis, as shown in Fig. 11, to determine the values of the constant coefficient, $c$, the elastic modulus of the tungsten wire, $E$, and the viscosity coefficient, $\eta$. The resulting values, with the determination coefficient of $R^2 = 0.9985$, are demonstrated in Table I.

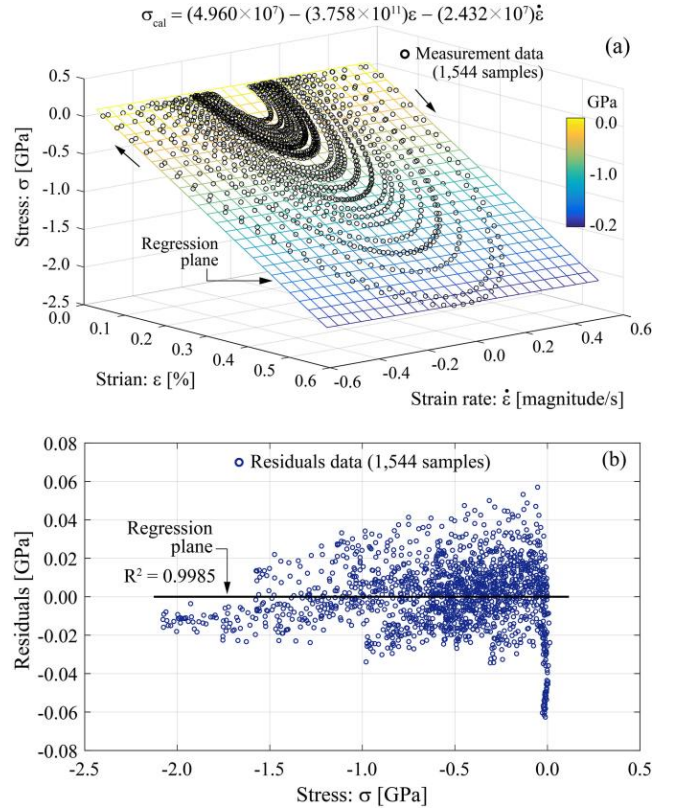

Fig. 11. Evaluation of the impact tensile stress estimation. (a) Multiple least-squares regression plane obtained by substituting the corresponding strain and strain rate into (5) with the appropriate coefficients $c$, $E$, and $\eta$ from Table I. (b) The correlation between the measured stress and its corresponding residuals.

TABLE I
COEFFICIENTS

| Quantities | Symbols | Values | Units |
|---|---|---|---|
| Constant coefficient | $c$ | $4.960 \times 10^7$ | N/m² |
| Elastic modulus | $E$ | $-3.758 \times 10^{11}$ | N/m² |
| Viscosity coefficient | $\eta$ | $-2.432 \times 10^7$ | N·s/m² |

The accuracy of the calculation method used to approximate some examples of the impact tensile stress, selected from the 30 experiments, was examined as shown in Fig. 12 and 13. The comparison between the impact tensile stress, obtained by measuring the inertial force acting on the moving mass, $\sigma_{mea}$, and by the corresponding calculation through the appropriate coefficients provided in Table I to (5), $\sigma_{cal}$, is shown in Fig. 12. The findings demonstrate that the calculated stress values are reasonably consistent with the corresponding measured values, indicating that the calculation method can be applicable for satisfactorily approximating the impact tensile stress.

The difference in stress values obtained by the measurement and the calculation, $\sigma_{mea}$ and $\sigma_{cal}$, is illustrated in Fig. 13. The root mean square (rms) of the difference values is 0.0096 GPa, 0.0086 GPa, 0.0126 GPa, and 0.0150 GPa for the varieties of selected experiments of 7th, 14th, 21st, and 29th, respectively. These difference values are corresponding to 1.75%, 1.16%, 1.19%, and 0.75% of the maximum impact tensile stress in each the experiment of approximately 0.5476 GPa, 0.7402 GPa, 1.0550 GPa, and 2.0008 GPa, respectively.

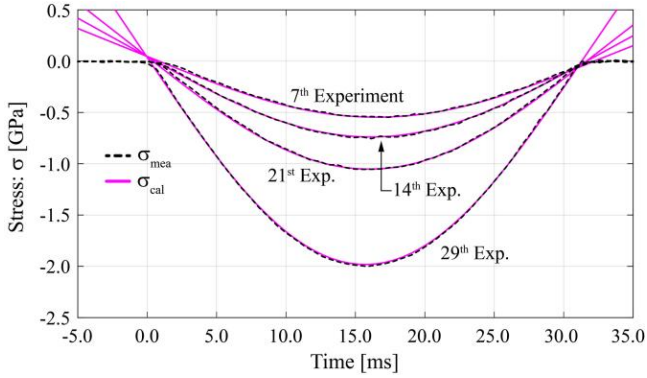

Fig. 12. Comparison between some examples of $\sigma_{mea}$ and $\sigma_{cal}$.

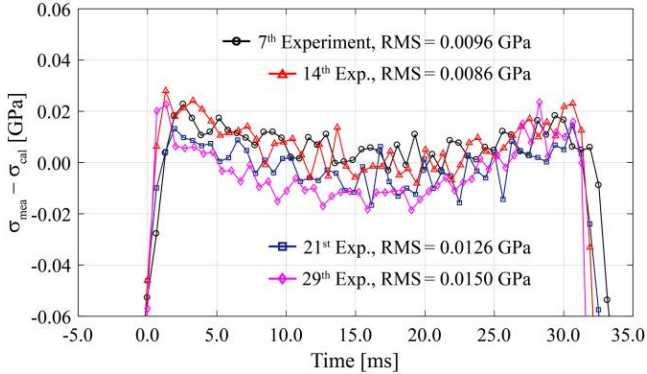

Fig. 13. Difference between the $\sigma_{mea}$ and $\sigma_{cal}$.

*B. Uncertainty evaluation*

The determination of uncertainty components in the measurement of the tensile force acting on the tungsten wire is carried out as follows:

U1. *Mechanical vibration of the optical interferometer:* The uncertainty due to mechanical vibration in the optical interferometer was evaluated by calculating the quadratic mean deviation (RMS value of the standard deviation) of the velocity, both before and after the impact tensile force, $F$, begins to take place on the tungsten wire under test. Thus, 20 data points were collected for each the velocity measurement. Based on the analysis of 30 measurements, the quadratic mean deviation of the velocity was approximately $3.29 \times 10^{-5}$ m/s and $3.01 \times 10^{-5}$ m/s, for the velocity before and after the impact tensile force begins to take place, respectively. Thus, it can be inferred that the variation in the velocity is approximately $3.29 \times 10^{-5}$ m/s, which corresponds to the acceleration variation of $1.15 \times 10^{-2}$ m/s² and the force variation of 33.5 mN.

U2. *Mass calibration:* The entire mass of the moving components was measured with the aid of an electric balance, which typically yielded a standard uncertainty of approximately 0.01 g, corresponding to 0.00035% of the total mass of the moving mass components, 2.897 kg. Resulting in the impact tensile force when it reaches a maximum value of approximately 16.30 N.

U3. *Optical alignment:* The uncertainty in the optical alignment was estimated by considering the inclination of the signal beam, which is approximately $1 \times 10^{-3}$ rad. This inclination angle of the signal beam resulted in a relative uncertainty in the velocity of approximately $5 \times 10^{-7}$ m/s, which is negligible.

U4. *Frequency stability:* The frequency stability of a Zeeman-type, two-wavelength He-Ne laser was estimated to have an uncertainty of 10 Hz. This uncertainty results in an approximate uncertainty of $3 \times 10^{-6}$ m/s in the velocity of the moving mass, corresponding to a negligibly small uncertainty of approximately 2.8 mN in the impact tensile force measurement.

U5. *Frictional force of the linear bearing:* The uncertainty associated with the frictional force acting within the aerostatic linear bearing under the condition where the air film remains unbroken and of which the thickness is approximately 8 μm was estimated using the method presented in [14]. The dynamic frictional force was calculated using the following equation:

$$F_{df} = Av \quad (6)$$

where $F_{df}$ represents the dynamic frictional force acting on the moving mass, $v$ represents its velocity, and $A = 8 \times 10^{-2}$ N·s/m is the constant coefficient of the bearing. The dynamic frictional force was calculated to be approximately 4.8 mN at a velocity of approximately 0.06 m/s. Given its small value, the dynamic frictional force can thus be neglected.

8Consequently, the standard uncertainty associated with measuring the impact tensile force applied to the tungsten wire was estimated to be 33.5 mN. Corresponding to 0.2% of the maximum impact tensile force, $F_{max}$, which was approximately 16.30 N.

In each experiment, the observation of the tungsten wire's behavior under impact tensile testing was brief, i.e., within 50 ms, as demonstrated in Fig. 6. This brief period allowed us to determine that an interval time of 0.5 s for the digitizer was sufficient. Based on 30 measurements, it was found that the kinetic energy dissipation was obtained in a small amount, with a maximum value of approximately 0.12 mJ when the maximum kinetic energy was estimated to be 4.50 mJ. The proposed method was found to be reproducibility, as the linear portions of all curves were consistent with each other, as shown in Fig. 7.

The proposed method allows for an easy setup procedure for testing various materials. The clamps allow for any type of wire to be attached, and the impact tensile force applied to the material can be evaluated with a high accuracy employing an optical interferometer.

During the experiments, only the time-varying beat frequency caused by motion was measured using the proposed method. Subsequently, the velocity, displacement, acceleration, and force were calculated based on the measured beat frequency. This approach allowed for the quantities to be obtained in a synchronized manner. Additionally, in accordance with the law of action-reaction, the impact tensile force exerted on the tungsten wire was determined by calculating the dynamic force acting on the moving mass, under the assumption that the friction force of the aerostatic linear bearing can be neglected. The dynamic force was directly measured using the definition of force as mass multiplied by acceleration. Furthermore, the uncertainty of the impact tensile force can be accurately evaluated from the uncertainties of both the mass and the acceleration.

## V. CONCLUSION

In this study, we propose an optical method for evaluating the mechanical properties of a tungsten wire under impact tensile load. In the method, an impact tensile load was applied to the wire under test employing the inertial force of a rigid mass supported by an aerostatic linear bearing. During the experiment, the velocity, displacement, acceleration, and force acting on the moving mass were observed using an optical interferometer. To attest the effectiveness of the proposed method, 30 measurements of the impact tensile force applied on the tungsten wire were performed. The results show that the amount of kinetic energy dissipation was low, indicating that the tungsten wire was operating within its elastic limit and exhibiting good recoverability, as it deforms under impact tensile load and then quite completely returns to its original shape. Moreover, the 30 experimental curves examined in this investigation displayed consistent linear portions, indicating the reproducibility of the proposed method. The results of the study offer valuable insights into the behavior of wires under dynamic impact loading, providing a deeper understanding of their mechanical properties and potential applications. These will be of interest to researchers and professionals in various fields such as material science, optical metrology, measurement science, and engineering.